\documentclass{cimento}

\usepackage{graphicx} 
\usepackage{float}

\title{The gravitational field of isolated objects in quadratic gravity}
\author{S.~Silveravalle
\thanks{In collaboration with Alfio Bonanno} 
\thanks{E-mail: samuele.silveravalle@unitn.it}
}
\instlist{\inst{} Universit\`a degli Studi di Trento, Via Sommarive, 14, IT-38123, Trento, Italy
  \inst{} TIFPA, Via Sommarive, 14, IT-38123, Trento, Italy}

\begin{document}

\maketitle

\begin{abstract}
When describing gravity at high energies it is natural to introduce terms quadratic in the curvature as first corrections to the Einstein-Hilbert action. Static, spherically symmetric classical solutions are studied in the case of the quantum-motivated $R+R^2+C^{\mu\nu\rho\sigma}C_{\mu\nu\rho\sigma}$ theory. In contrast with the case of General Relativity, where the metric is forced to be the Schwarzschild one, a large spectrum of solutions with different gravitational potentials has been found. It is shown how these solutions populate the parameter space of the theory, and some possible phenomenological implications of this modified potential are discussed.
\end{abstract}

\section{Introduction}

Gravity has always been the guiding light of Physics into field theories; nevertheless, it is the only fundamental interaction for which there is no general consensus on a Quantum Field Theory description. In a standard, perturbative QFT context, General Relativity can be consistent only as a low energy effective theory, having divergent corrections at higher energies. The form of the counterterms at one-loop has made clear that the first sensible corrections to the Einstein-Hilbert action are terms quadratic in the curvature \cite{thooft}. Such terms appear also in the low energy limit of String Theory \cite{zwiebach}, and in non perturbative analysis of the Renormalisation Group flow of gravity \cite{benedetti}. The most general quadratic Lagrangian in four dimension without cosmological constant can be written as
\begin{equation}\label{qga}
\mathcal{S}_{QG}=\int \drm^4x\, \sqrt{-g}\left[\gamma\, R+\beta\, R^2-\alpha\, C_{\mu\nu\rho\sigma}C^{\mu\nu\rho\sigma}\right],
\end{equation}
where $R$ is the Ricci scalar and $C_{\mu\nu\rho\sigma}$ is the Weyl tensor. Within this theory gravity has three mediators: the standard massless graviton, a massive scalar, and a massive spin-2 particle. At quantum level it has been proven by K. Stelle that this theory is renormalisable, but at the price of a loss of unitarity due to the presence of negative energy states for the spin-2 particle \cite{Stelle1}. While there have been various proposals to solve this issue \cite{piva,donog,platania}, a more conservative approach, where the action is not considered as a full quantum theory but only as a first-order correction of an unknown theory of quantum gravity, has been preferred. Considering the major contribution of quadratic curvature terms in the description of the very early universe \cite{lehners}, the classical solutions of (\ref{qga}) are indeed good candidates to be first-order quantum corrections to the solutions of GR.\\
Solutions representing the most simple isolated objects, \textit{i.e.} static and spherically symmetric, have been studied extensively in recent times, both in the vacuum \cite{perkins,pravda,holdom,Alfio1} and the non-vacuum \cite{Alfio2} cases. In addition to the ones of General Relativity, many solutions have been found: non-Schwarzschild black holes, wormholes and exotic naked singularities. This large variety in the spectrum of solutions is in contrast with the classical case, where the metric is forced to be the one of Schwarzschild, and is completely characterised by the total mass $M$. In this communication it is shown how the gravitational potential is modified by the different types of solutions, picturing a sort of ``phase diagram'' of the theory. Moreover, it is shown how these modifications can have relevant implications for astrophysical observations of compact stars \cite{Alfio2}. 

\section{Analytical and numerical setups}

\subsection{Equations of motion and analytical approximations}

As the ansatz for the metric it has been chosen the static, spherically symmetric one in Schwarzschild coordinates
\begin{equation}\label{met}
\drm s^2=-h(r)\,\drm t^2+\frac{\drm r^2}{f(r)}+r^2\drm \Omega^2,
\end{equation}
and, in the case of compact stars, the perfect fluid stress-energy tensor which, with these symmetries, depends only from the energy density $\rho(r)$ and isotropic pressure $p(r)$. From the minimisation of the action and the conservation of the stress-energy tensor it is possible to derive a set of three ordinary differential equations at third order in $h(r)$ and $f(r)$, and first order in $\rho(r)$ and $p(r)$. The introduction of an equation of state $p=\mathcal{P}\left(\rho\right)$ specifies the physical nature of the fluid and closes the system of equations.\\
Considering isolated objects, at large distances it is possible to express the metric as a small perturbation around the flat one and linearise the equations. As shown in \cite{perkins,Alfio1} the solution for the metric in this weak field limit is
\begin{eqnletter}\label{wfm}
h(r) & = & 1-\frac{2M}{r}+2S_2^-\frac{\mathrm{e}^{-m_2\,r}}{r}+S_0^-\frac{\mathrm{e}^{-m_0\,r}}{r},\label{wfh}\\
f(r) & = & 1-\frac{2M}{r}+S_2^-\frac{\mathrm{e}^{-m_2\,r}}{r}\left(1+m_2\,r\right)-S_0^-\frac{\mathrm{e}^{-m_0\,r}}{r}\left(1+m_0\,r\right),\label{wff}
\end{eqnletter} 
where the masses are $m_0^2=\gamma/6\beta$ and $m_2^2=\gamma/2\alpha$, and are respectively the ones of the scalar and spin-2 particles of the quantum theory. From (\ref{wfh}) it can be seen that the gravitational potential $\phi(r)=\frac{1}{2}\left(h(r)-1\right)$ is not completely determined by the total mass $M$, but also by two ``charges'' $S_2^-$ and $S_0^-$ associated with Yukawa corrections.\\
The different types of solutions are classified by the first exponents of a Frobenius-like series expansion around either the origin or a radius $r_0\neq 0$ \cite{perkins,pravda}. In particular, black holes appear when the functions $h(r)$ and $f(r)$ go to zero linearly at $r=r_H$, and compact stars when there is no horizon and the metric is regular everywhere. There are, however, solutions that are not present in GR; understanding in which way the parameters of the gravitational potential $(M,S_2^-,S_0^-)$ relate with the different families of solutions is the major point addressed in this communication.

\subsection{Numerical procedures}

The search for solutions of (\ref{qga}) in this setting is reduced to solving a system of third order ODEs. Two different approaches have been used:
\begin{itemize}
\item[-] an \textit{a posteriori} classification, where the equations are integrated from a large radii with initial values given by (\ref{wfm}) and classified by their behaviour at a radius $r_0\neq 0$ or at the origin;
\item[-] an \textit{a priori} characterization, where a type of solution is chosen by fixing the metric at an internal boundary and is matched with the weak field limit at large radii with a shooting method technique.
\end{itemize}
The first method has been used to study the phase diagram of the theory, while the second one has proven to be efficient to study specific types of solutions (a more detail explanation on the use of the shooting method can be found in \cite{Alfio1,Alfio2}).

\section{Isolated objects in quadratic gravity}

\begin{figure}[H]
\begin{center}
\includegraphics[width=0.85\textwidth]{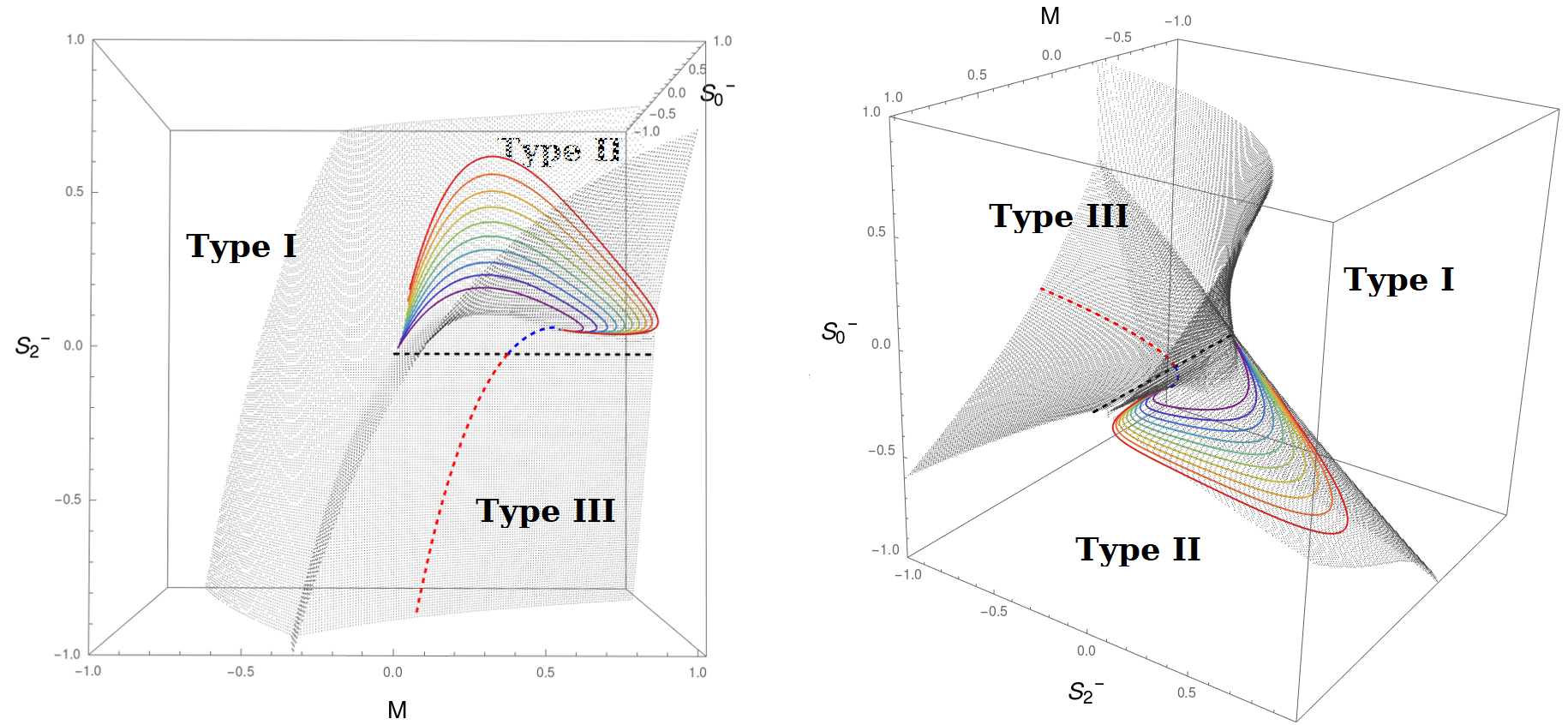}
\caption{Parameter space of quadratic gravity for $m_0/m_2=1.1$. Type I, II and III are exotic solutions, the black dashed line indicates Schwarzschild black holes while the dashed red and blue line indicate non-Schwarzschild ones, solid lines instead indicate compact stars solutions with different colors standing for different values of $k_0$ in a polytropic e.o.s. $p=k_0\rho^2$.}\label{pd}
\vspace*{-0.5cm}
\end{center}
\end{figure}

In fig. \ref{pd} the phase diagram of the solutions is presented. There are three macro-regions populated by exotic solutions not present in General Relativity, while the familiar black hole and compact star solutions are present only in zero-measure lines, lying on the surfaces that separate the macro-regions.\\[-0.6cm]
\noindent
The three types of exotic solutions are respectively naked singularities with a divergent metric, naked singularities with a vanishing metric (discussed in detail in \cite{holdom}) and non-symmetric wormholes. These latter solutions in particular are good candidates to be black hole mimickers and they populate a large volume of the physical part of the phase diagram (\textit{i.e.} with $M>0$). A complete description of the phase diagram of the solutions will be given soon in a longer paper.\\[-0.5cm]

\begin{figure}[h]
\begin{center}
\includegraphics[width=0.85\textwidth]{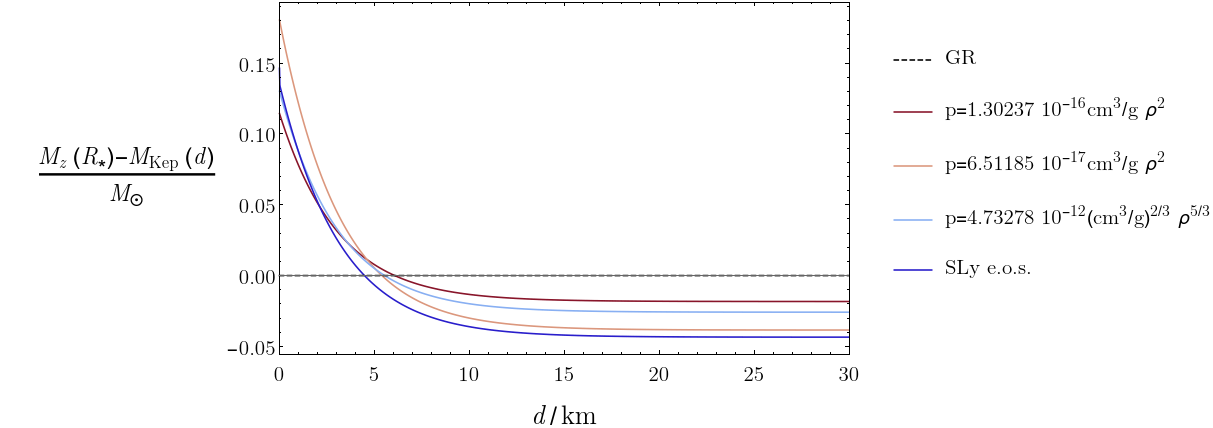}
\caption{Differences between masses measured with surface redshift or planetary transit, in function of the distance from the star surface, for stars with different equations of state.
\vspace*{-1cm}}\label{eoscs}
\end{center}
\end{figure}

At last, it is interesting to focus on the consequences of having different corrections to the gravitational potential with a toy-model thought experiment. A neutron star with a satellite orbiting at a distance $d$ from the surface is considered; in principle it is possible to measure the neutron star mass both from the redshift of a photon emitted at the star surface $M_{z}\left(R_*\right)$, and from the transit of the satellite with the use of Kepler's third law $M_{Kep}\left(d\right)$. In General Relativity, all possible mass definitions coincide with the Schwarzschild parameter $M$, and the difference from these two measurements would be zero. In quadratic gravity, instead, not only there would be a difference, but such difference would be sensitive to the precise nature of the solution, as the equation of state of the fluid, as shown in fig. \ref{eoscs} for stars with either a polytropic or the realistic SLy e.o.s.\\[-0.07cm]
 
\section{Conclusions}

Quadratic gravity is one of the most simple and natural extensions of General Relativity at high energies. Nevertheless, even in the most simplified setting, the spectrum of possible solutions is much richer than in the classical case. The solutions of General Relativity are present only in zero-measure regions of the parameter space, which is mainly populated by exotic solutions. It is still to determine whether all these exotic solutions are present also in the physical spectrum of the theory. However, the Yukawa corrections to the gravitational potential suggest a possible test of General Relativity, and a rich phenomenology that still has to be considered in a realistic scenario.

\end{document}